\documentclass[11pt, a4paper]{article}

\usepackage {fullpage}						
\usepackage [bookmarks=false, colorlinks=true, linkcolor=blue, citecolor=blue, urlcolor=blue]{hyperref} 

\usepackage{multimedia}
\usepackage {color}

\usepackage {ucs}
\usepackage{upgreek} 						
\usepackage [utf8x]{inputenc}
\usepackage {graphicx}

\usepackage {amsfonts,amsmath,amsthm,amssymb}
\usepackage {latexsym, bbm}
\usepackage {graphics, graphicx}
\usepackage {fancyhdr}
\usepackage[table]{xcolor}
\usepackage {epstopdf} 						

\usepackage {multicol}						
\usepackage {multirow}  					
\usepackage {soul} 							

\usepackage{courier}						

\usepackage {authblk} 						
\usepackage {datetime} 						
\usepackage {subfigure}						
\usepackage {parskip}						
\usepackage {setspace}						
\usepackage[font=small,textfont=it,width=0.97\textwidth]{caption}		
\usepackage {mathtools}						
\usepackage {array}							
\usepackage{tabularx}

\renewcommand {\vec}{\mathbf}

\newcommand{\snodes}{s_\text{nodes}}
\newcommand{\anodes}{\alpha_\text{nodes}}
\renewcommand{\a}{\alpha}

\theoremstyle{definition}
\newtheorem{experiment}{Experiment}[section]

\renewcommand{\(}{\left(}
\renewcommand{\)}{\right)}
\newcommand{\R}{\mathbb{R}}

\renewcommand{\P}{\mathbb{P}}

\graphicspath{{./}}

\def\gray#1{{\color{gray}\text{#1}}}

\begin{document}
\title{Chemotaxis and Haptotaxis on Cellular Level}
\author[1]{A. Brunk}
\author[2]{N. Kolbe}
\author[3]{N. Sfakianakis} 

\affil[1]{{\footnotesize aabrunk@students.uni-mainz.de, Institute of Mathematics, Johannes Gutenberg-University}}
\affil[2]{{\footnotesize kolbe@uni-mainz.de, Institute of Mathematics, Johannes Gutenberg-University}}
\affil[3]{{\footnotesize sfakiana@uni-mainz.de, Institute of Mathematics, Johannes Gutenberg-University}}

\maketitle

\begin{abstract}
Chemotaxis and  haptotaxis have been a main theme in the macroscopic study of bacterial and cellular motility. In this work we investigate the influence these processes have on the shape and motility of fast migrating cells. We note that despite their biological and modelling differences, the cells exhibit many similarities in their migration. We moreover see, that after an initial adjustment phase, the cells obtain a stable shape and their motion is similar to advection. 
\end{abstract}

\section{Introduction} 

In the biology of many diseases and in particular, in the \textit{growth} and \textit{metastasis} of \textit{cancer}, the biological processes of \textit{chemotaxis} and \textit{haptotaxis} play a fundamental role, see e.g. \cite{Condeelis.2011}.

Chemotaxis is the directed motion of biological organisms (bacteria, cells, multicellular organisms) as response to an extracellular chemical signal. Due to their size, cells can identify spatial gradients of the chemical ingredient on their membrane and adjust their migration accordingly.

Haptotaxis, on the other hand, is the directed cell motion as response to a gradient of extracellular adhesion sites or substrate-bound chemo-attractant/repellents. The cells attach on the adhesion sites by use of specialized transmembrane proteins like the \textit{integrins}. They are an indispensable part of the motility apparatus of the cells.

From a mathematical point of view they are most often studied in the spirit of Keller-Segel systems \cite{KS.1970, Chaplain.2007, Sfak-Kol-Hel-Luk.2016,Kol-Kat-Sfak-Hel-Luk.2014}. In such approaches, the involved quantities are represented by their \textit{macroscopic} density. 

This has created a gap between the mathematical investigations ---at least in the macroscopic approach--- and the experimental biological/medical sciences where most of the knowledge/understanding refers to single cells and their properties. The current work is an effort to shed some light in this research direction. 

\begin{figure}[t]
	\centering
		\subfigure[]{{\includegraphics[height=10em]{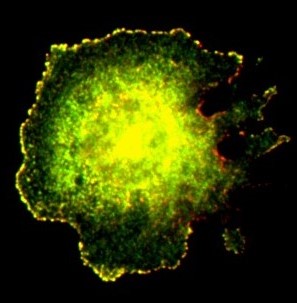}}}\hspace{3em}
		\subfigure[]{{\includegraphics[height=10em]{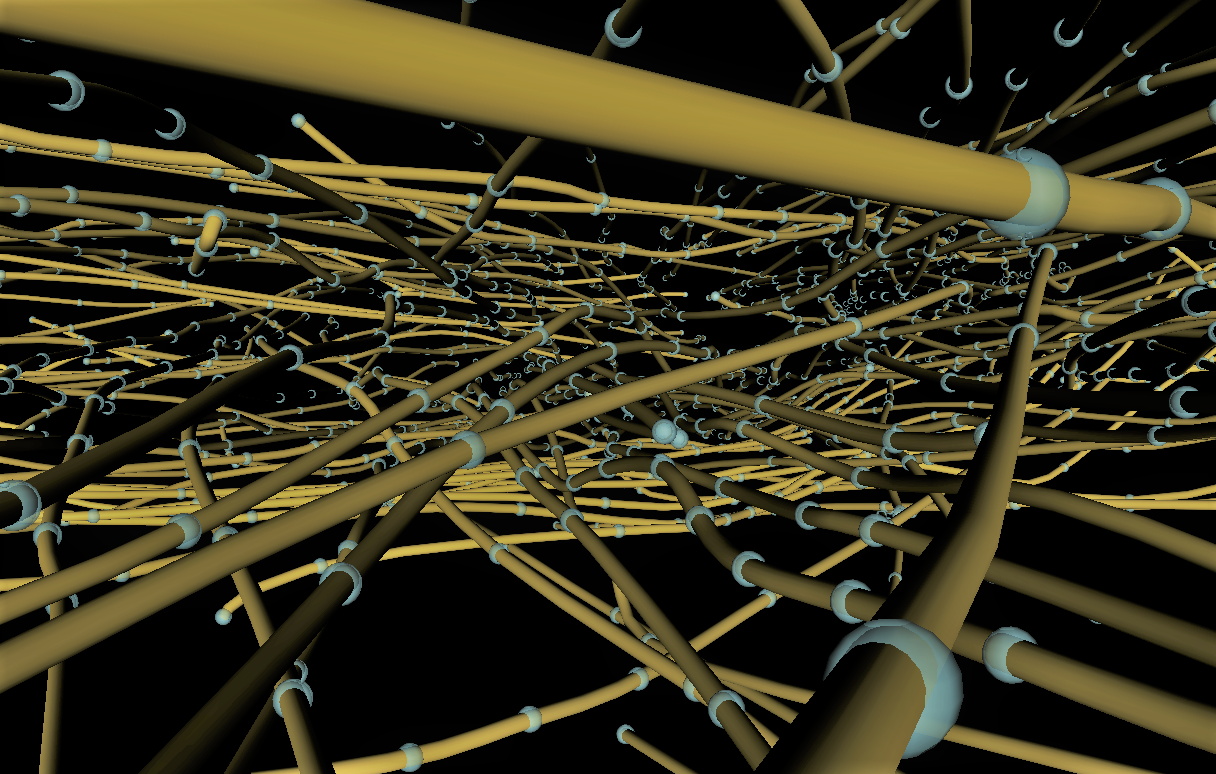}}}
	\caption{(a) NIH3T3 cell during migration. The lamellipodium is located in the light-coloured front of the cell; License:\textit{Cell Image Library, CIL:26542}.  (b) The inside of the lamellipodium as comprised by a large number of actin  filaments. Our reconstruction is based on experimental (blue spheres) data by Vic Small.}\label{fig:lamellip}
\end{figure}

In particular, we consider very motile cells like \textit{fibroblast}, \textit{keratocyte}, or even cancer cells, that migrate over adhesive substrates. These cells develop thin protrusions, called \textit{lamellipodia} (singular \textit{lamellipodium}), see Fig. \ref{fig:lamellip} and \cite{Small1978}. They can be found at the leading edge of the cells, and are comprised of a network of \textit{actin-filaments} which are highly dynamic linear \textit{bio-polymers} \cite{Small2002, Lauffenburger1996}. Intra- or extracellular reasons might lead to polarizations of the lamellipodium and to cell motion similar to ``crawling''  \cite{Svitkina1997, Iijima2002}.

We follow here the modelling approach introduced in \cite{Oelz2008} and extended in \cite{MOSS-model}, and consider the \textit{Filament Based Lamellipodium Model} (FBLM); a two-dimensional continuum model that describes the dynamics of the actin meshwork  and results to the motility of the cell. It distinguishes between two different families of filaments and takes into account the interactions between them and the \textit{Extracellular Matrix} (ECM). Numerically, we have developed a problem specific \textit{Finite Element Method} (FEM) that allows for efficient investigation of the FBLM. Fore more details see  \cite{OelzSch-JMB, Schmeiser2010, MOSS-numeric}.  

We investigate comparatively the influence that chemotaxis and haptotaxis have on the shape and the motion of migrating cell. In more detail: in section \ref{sec:FBLM} we present the main components of the FBLM and address the FEM that we use to numerically resolve the FBLM. In Sections \ref{sec:chemo} and \ref{sec:hapt} we elucidate the way that chemotaxis and haptotaxis are incorporated in our study and what the ensuing cell motility looks like, and in Section \ref{sec:comparison} we present comparative results between the two cases.

\section{Mathematical model and numerical method} \label{sec:FBLM}

This section is devoted to the brief presentation of the FBLM and the FEM that we use to numerically solve it.

\subsection{The FBLM} 
\begin{figure}[t]
	\centering
	\includegraphics[height=13em]{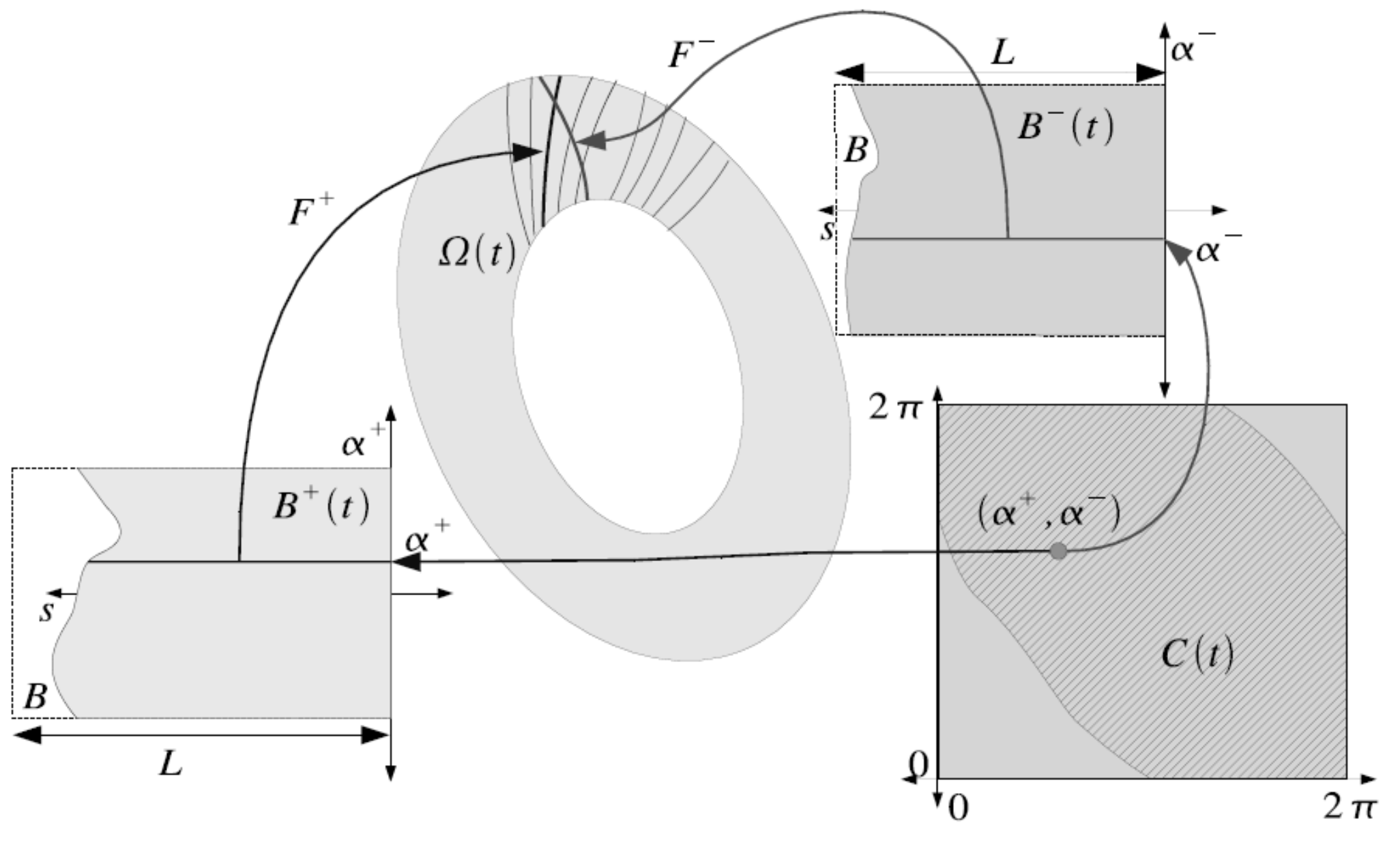}
	\caption{Graphical representation of (\ref{eq:strong}); showing here the lamellipodium $\Omega(t)$ as a mapping of $\vec{F}^\pm$ and the crossing--filament domain $\mathcal C$, see also \cite{Schmeiser2010}.}\label{fig:domains}
\end{figure}

We present here the main components of the FBLM and refer to \cite{Oelz2008,MOSS-model,Schmeiser2010} for details.  The first assumption behind the model is that the lamellipodium is a two dimensional structure comprised of actin filaments organized in two locally parallel families, see Fig. \ref{fig:domains}.

The filaments of each family (denoted as $\vec F^\pm$) are labelled by an index $\alpha\in [0,2\pi)$,  they have length $L^\pm(\alpha,t)$ at time $t$, and can be parametrized with respect to its arclength as
$\left\{\vec{F}^\pm(\alpha,s,t): -L^\pm (\alpha,t)\le s \le 0\right\}\subset {\R}^2,$ where the membrane corresponds to $s=0$. 

The two families define identical outer boundaries 
\begin{equation}\label{eq:tether}
	\left\{\vec{F}^+(\alpha,0,t): 0\le \alpha < 2\pi\right\} = \left\{\vec{F}^-(\alpha,0,t): 0\le \alpha < 2\pi\right\} \,,
\end{equation}
which, along with the \emph{inextensibility} assumption along the filaments:
\begin{equation}\label{eq:inext}
	\left|\partial_s \vec{F}^\pm(\alpha,s,t)\right| = 1 \qquad\forall\, (\alpha,s,t) \;,
\end{equation}
constitute additional constraints for the unknowns $\vec{F}^\pm$.

The FBLM reads for the family $\vec F=\vec F^+$ (symmetrically for $\vec F=\vec F^-$) as:
\begin{eqnarray} 
	0&=&\underbrace{\mu^A \eta D_t \vec{F}}_\text{\gray{adhesion}} -\underbrace{\partial_s\left(\eta \lambda_{\rm inext} \partial_s \vec{F}\right)}_\text{\gray{in-extensibility}}  + 	\underbrace{\partial_s\left( p(\rho) \partial_\a \vec{F}^{\perp}\right)
	-\partial_\a \left( p(\rho) \partial_s \vec{F}^{\perp}\right)}_\text{\gray{pressure}} \nonumber\\
	&+& \underbrace{\mu^B \partial_s^2\left(\eta \partial_s^2 \vec{F}\right)}_\text{\gray{bending}}\pm\underbrace{\partial_s\left(\eta\eta^- \widehat{\mu^T} (\phi-\phi_0)\partial_s \vec{F}^{\perp}\right)}_\text{\gray{twisting}}  
	+ \underbrace{\eta\eta^- \widehat{\mu^S}\left(D_t \vec{F} - D_t^- \vec{F}^\mp\right)}_\text{\gray{stretching}}\,, ~~\label{eq:strong}
\end{eqnarray}
with $\vec{F}^\bot = (F_1,F_2)^\bot = (-F_2,F_1)$. The function  $\eta(\alpha,s,t)$ represents the number density of filaments of the family $\vec F$  length at least $-s$ at time $t$ with respect to $\alpha$. 

The first term in the model is responsible for the interaction of the intra- with the extracellular environment, and in particular for the momentum transfer between the cell and the ECM and for the motility of the cell. It is the prominent term of \eqref{eq:strong} and, in this sense, the model \eqref{eq:strong} is advection dominated. 
	
The polymerization speed of the filaments is given by $v(\alpha,t)\ge 0$, and the \textit{material derivative} 
\begin{equation}
	D_t \vec{F} := \partial_t \vec{F} - v\partial_s \vec{F},
\end{equation}
describes the velocity of the actin material relative to the substrate. 
 
The inextensibility term follows from the constraint (\ref{eq:inext}) with a \textit{Lagrange multiplier} $\lambda_{\rm inext}(\alpha,s,t)$. The pressure term models the electrostatic repulsion between filaments of the same family. The filament bending is modelled according to Kirchhoffs bending theory. The last two terms in (\ref{eq:strong}) model the interaction between the two families caused by \textit{elastic cross-link} junctions. They resist against twisting away from the equilibrium angle $\phi_0$ of the cross-linking molecule, and against the stretching between filaments of the two families.

The system (\ref{eq:strong}) is subject to the boundary conditions 
\begin{eqnarray} \label{eq:newBC}
	- \mu^B\partial_s\left(\eta\partial_s^2 \vec{F}\right) &- &p(\rho)\partial_\a \vec{F}^\perp + \eta \lambda_{\rm inext} \partial_s \vec{F}  
		\mp\eta\eta^* \widehat{\mu^T}(\phi-\phi_0)\partial_s \vec{F}^\perp \\
		&=&\left\{
			\begin{array}{l l}
				\eta \left(f_{\rm tan}(\alpha)\partial_s \vec{F} + f_{\rm inn}(\alpha) \vec{V}(\alpha)\right), & \quad \mbox{for 	} s=-L \;,\\
				\pm\lambda_{\rm tether} \nu, & \quad  \mbox{for }  s=0 \;,
			\end{array}\right.\nonumber\\
		\eta \partial_s^2 \vec{F} &= &0, \quad \mbox{for } s=-L,0 \;,\nonumber
\end{eqnarray}
where at $s=0$ they describe the forces due to the tethering of the filaments to the membrane, and at $s=-L$ the contraction effect of the actin-myosin interaction with the interior region, see \cite{MOSS-model} for details.



\subsection{The FEM} 

For the numerical treatment of the FBLM, we have developed a problem specific FEM. We present here its main characteristics and refer to \cite{MOSS-numeric} for details.

The maximal filament length varies around the lamellipodium, hence, the computational domain 
\[
	B(t) = \left\{(\alpha,s):\, 0\le\a<2\pi\,,\, -L(\alpha,t)\le s < 0\right\}
\]
is non-rectangular, see Fig. \ref{fig:domains}.  The orthogonality of the domain is recovered using the particular coordinate transformation $(\alpha,s,t) \rightarrow \(\alpha, L(\alpha,t)s, t\)$, and gives rise to the new domain $B_0 := [0,2\pi)\times[-1,0)\ni (\alpha,s)$. 

We discretize $B_0$ as $B_0 = \bigcup_{i=1}^{N_a} \bigcup_{j=1}^{N_s -1} C_{i,j}$, 
with $C_{i,j}=[\alpha_i,\alpha_{i+1})\times[s_j,s_{j+1})$ where $\alpha_i = (i-1)\Delta \alpha$, $\Delta \alpha = \frac{2\pi}{N_\a}$, and $s_j = -1 + (j-1)\Delta s$, $\Delta s = \frac{1}{N_s - 1}$.

The conforming Finite Element space we consider is
\begin{eqnarray}
\mathcal V := \Bigl\{ &\vec{F}&\in C_\a\([0,2\pi];\, C^1_s([-1,0])\)^2 \text{ such that } \vec{F}\bigm|_{C_{i,j}}(\cdot,s) \in \P^1_\a\,,\, \nonumber \\
&\vec{F}&\bigm|_{C_{i,j}}(\a,\cdot) \in \P^3_s\quad\mbox{for } i=1,\ldots,N_\a\,;\,  j = 1,\ldots,N_s-1\Bigr\}  \,, \label{eq:FE:V}
\end{eqnarray}
and includes the, per direction and per cell $C_{i,j}$, shape functions
\begin{equation}\label{eq:H}
\left. \begin{array}{lcl}
H_k^{i,j}(\a,s)=L_1^{i,j}(\a) G_k^{i,j}(s), & \mbox{ for }k=1\dots4\\
H_k^{i,j}(\a,s)=L_2^{i,j}(\a) G_{k-4}^{i,j}(s), & \mbox{ for }k=5\dots8
\end{array}\right\},
\end{equation}	
where $L^{i,j}_{1...2}$ are linear functions of $\a$ and $G^{i,j}_{1...4}$ are cubic functions of $s$:
\begin{equation}\label{eq:LG}
\left.\begin{array}{lcl}
L_1^{i,j}(\a) =\frac{\a_{i+1}-\a}{\Delta \a}, 
&&\quad G_1^{i,j}(s)=1-\frac{3(s-s_j)^2}{\Delta s^2}+\frac{2(s-s_j)^3}{\Delta s^3} \\
L_2^{i,j}(\a)=1-L_1^{i,j}(\a),
&&\quad G_2^{i,j}(s)=s-s_j-\frac{2(s-s_j)^2}{\Delta s}+\frac{(s-s_j)^3}{\Delta s^2}\\
&&\quad G_3^{i,j}(s)=1-G_1^{i,j}(s)\\
&&\quad G_4^{i,j}(s)=-G_2^{i,j}(s_j+s_{j+1}-s)
\end{array}\right\},
\end{equation}
for $(\a,s)\in C_{i,j}$, and where $H_k^{i,j}(\a,s)=0$ for $(\a,s)\not\in C_{i,j}$.

Accordingly, the weak formulation of (\ref{eq:strong}) (neglecting the boundary conditions) reads as:
\begin{eqnarray}
0 &=&\int_{B_0}\eta \left( \mu^B \partial^2_s\vec{F} \cdot \partial_s^2\vec{G} + L^4\mu^A \widetilde{D_t} \vec{F} \cdot \vec{G} + L^2\lambda_\text{inext} \partial_s\vec{F} \cdot \partial_s\vec{G} \right) d(\alpha,s) \nonumber\\
&+&\int_{B_0}  \eta\eta^-\left( L^4\widehat{\mu^S} \(\widetilde{D_t}\vec{F}  - \widetilde{D_t^-}\vec{F}^- \)\cdot \vec{G} \mp L^2\widehat{\mu^T}(\phi-\phi_0)\partial_s\vec{F}^{\perp} \cdot \partial_s \vec{G} 
\right) \ d(\alpha,s) \nonumber\\
&-& \int_{B_0} p(\varrho) \left( L^3\partial_\a \vec{F}^{\perp}\cdot \partial_s \vec{G} - \frac{1}{L}\partial_s \vec{F}^{\perp}\cdot \partial_\a (L^4\vec{G})\right) d(\alpha,s) 
\,.\label{eq:FEM}
\end{eqnarray}
for $\vec{F}, \vec{G}\in H^1_\a\((0,2\pi);\,H^2_s(-1,0)\)$, and for the modified material derivative 
$$\widetilde{D_t} = \partial_t - \(\frac{v}{L} + \frac{s \partial_t L}{L}\)\partial_s$$
and in-extensibility constraint $\left|\partial_s \mathbf F(\alpha,s,t)\right| = L(\alpha,t)$.

\section{Chemotaxis driven cell migration}\label{sec:chemo}


\begin{figure}[t]
	\centering
	\begin{tabular}{cccc}
		{\includegraphics[height=8em]{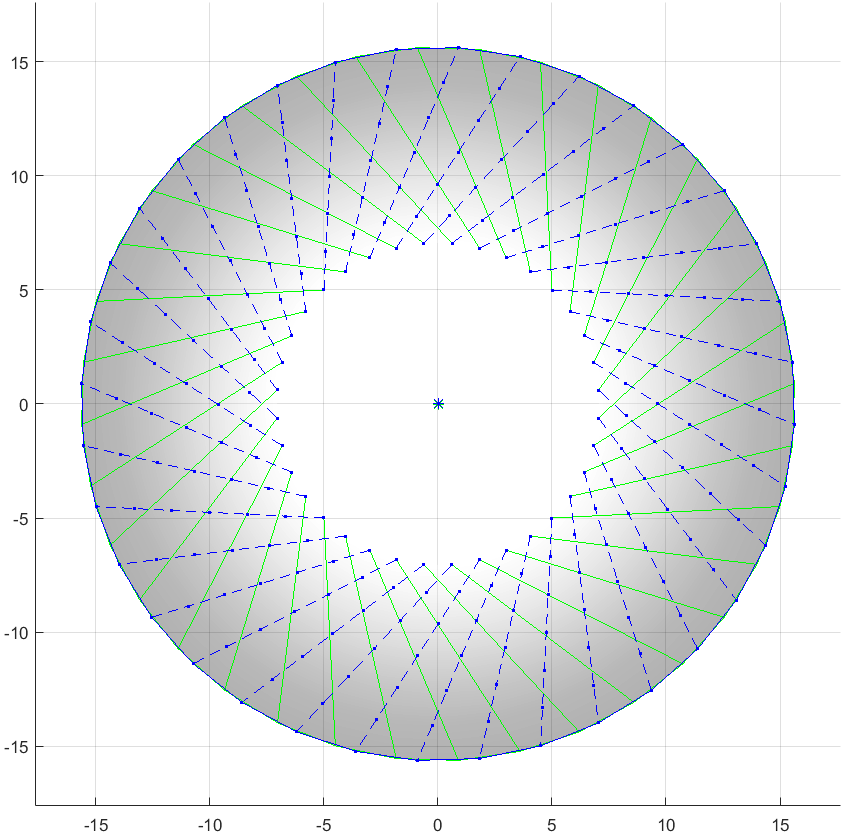}}&
		{\includegraphics[height=8em]{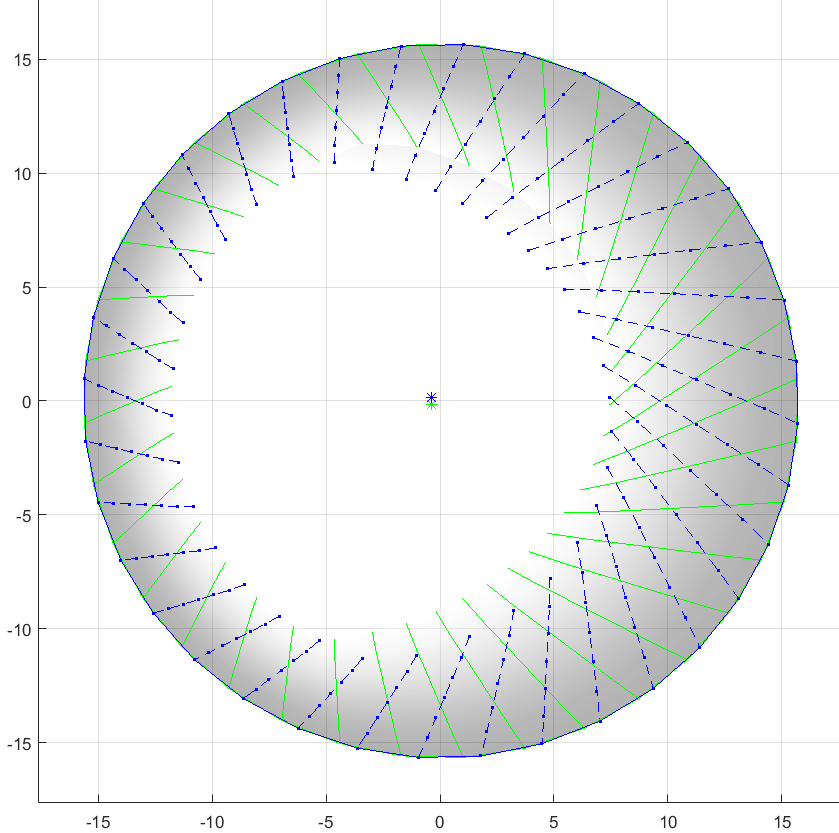}}&
		{\includegraphics[height=8em]{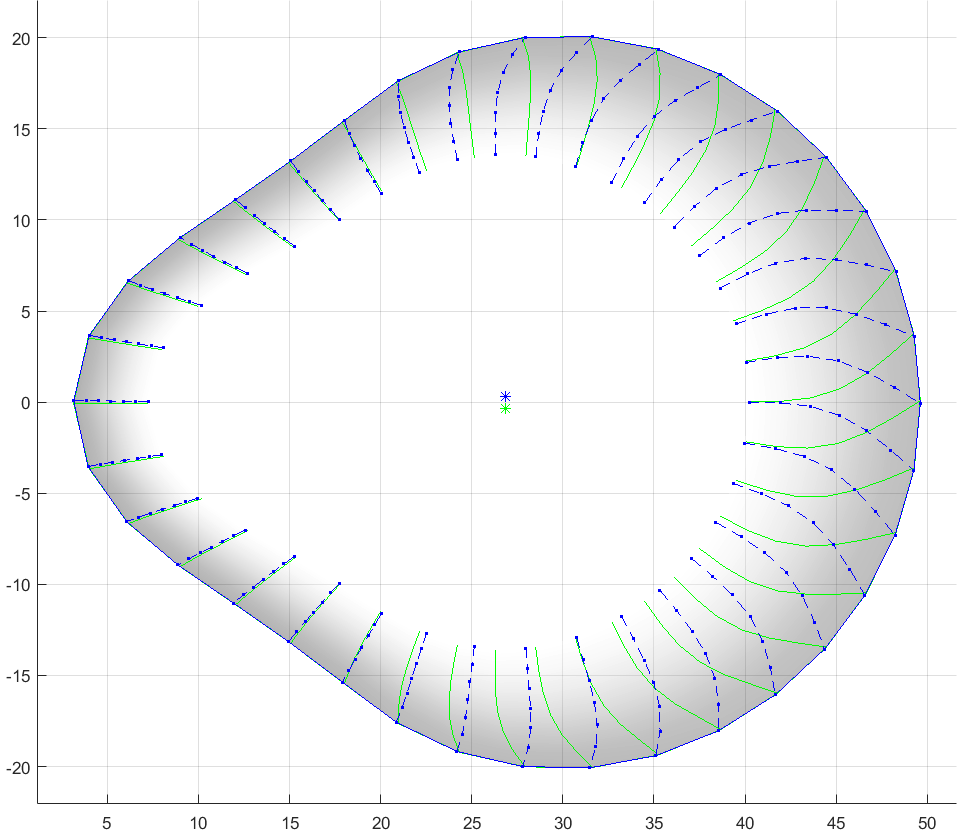}}&
		{\includegraphics[height=8em]{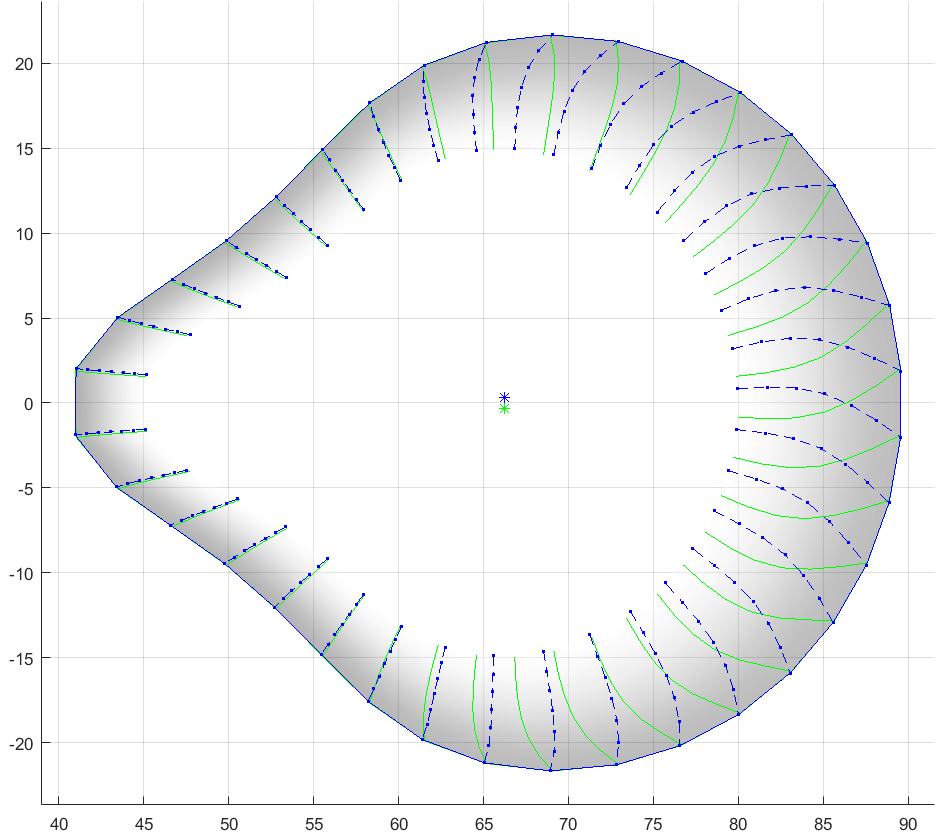}}\\
		$t=0.0015$&$t=0.0165$&$t=9.0015$&$t=18.4365$
	\end{tabular}
	\caption{Chemotaxis Experiment \ref{exp:ch.c.0.33} with $c=0.33$. The combination of the internal myosin retraction and the threshold value of $c$ give rise to the particular shape of the ``tail'' of the cell.}\label{fig:ch.c.0.33}
\end{figure}	

The sensing of the extracellular chemical component is prescribed directly on the membrane of the cell by the function
\[
	S(\a)=S_0 + S_1 \left( x\cos (\phi_{ca}) + y\sin(\phi_{ca}) \right),
\]
where $(x,y)$ lies on the membrane, $\phi_{ca}$ denotes the relative direction of the chemical signal with respect to the cell, and $S_0$ and $S_1$ the strength of the signal. We introduce a cut-off value $c$ on the relative signal intensity $S$ to account for the sensitivity of the cell to the low chemical ingredient densities. The higher the value of $c$ the smaller is the part of the cell that ``senses'' the chemical ingredient. Accordingly, the polymerization velocity $v_\text{ref}(\a)$ of every filament $\a$ is adjusted between two values $0<v_\text{min}<v_\text{max}$, as 
\begin{equation}
	v_\text{ref}(\a) = v_\text{min} + I(\a) \(v_\text{max}-v_\text{min}\),
\end{equation}
where $0\leq I(\a) =\frac{\max_{\a} S(\a) -S(\a)}{\max_\a S(\a) -\min_\a S(\a)} \leq 1$  represents the normalized response of the cell to the chemical ingredient, see also \cite{MOSS-model}.

Furthermore, the polymerization velocity $v$ is adjusted by the signed local curvature $\kappa$ of the membrane as
\begin{equation} \label{eq:v.mbrn} 
	v=\frac{2v_\text{ref}}{1+e^{\frac{\kappa}{\kappa_\text{ref}}}}
\end{equation}
where $\kappa_\text{ref}$ is the reference membrane curvature related to the local intensity of the chemical signal. The polymerization velocity $v$  influences also the length of the lamellipodium as is adjusted by the formula:
\[
	L= - \frac{\kappa_\text{cap,eff}}{\kappa_\text{sev}} +  \sqrt{\frac{\kappa^2_\text{cap,eff}}{\kappa^2_\text{sev}} + \frac{2 v}{\kappa_\text{sev}}\log{ \frac{\eta(s=0)}{\eta_\text{min}} }},
\]
cf. with Table \ref{tbl:parameters} and \cite{MOSS-numeric} for the biological meaning of $\eta_\text{min}$. 

\begin{figure}[t]
	\centering
	\begin{tabular}{ccc}
		{\includegraphics[height=8em]{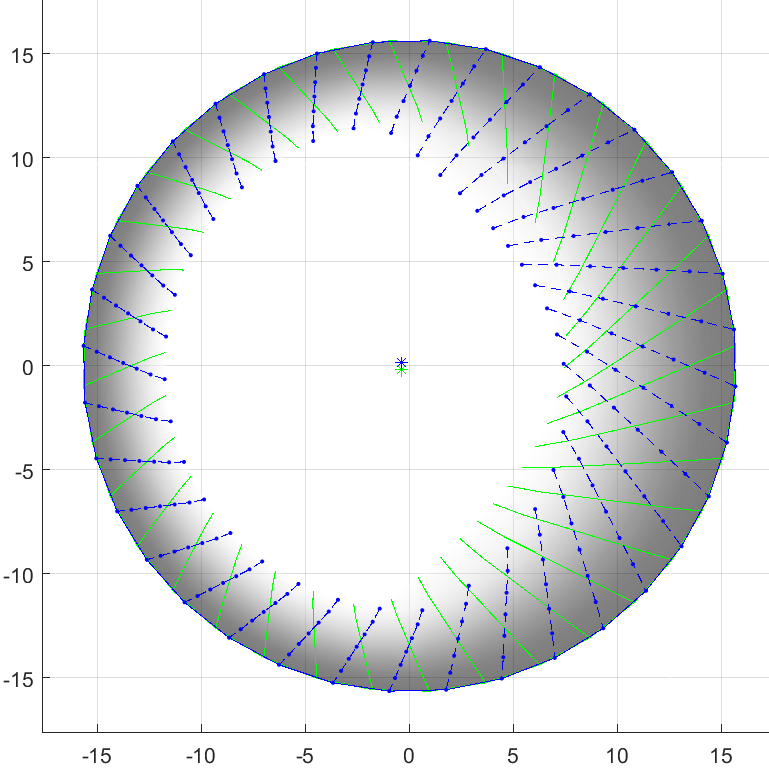}}&
		{\includegraphics[height=8em]{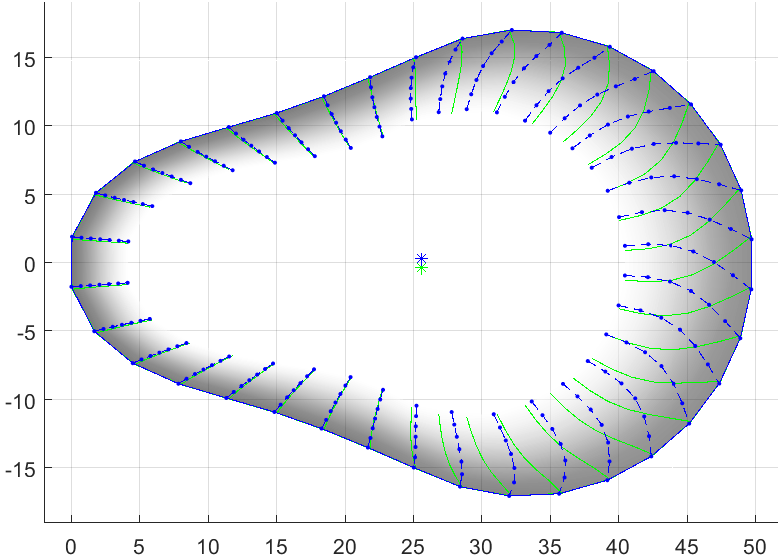}}&
		{\includegraphics[height=8em]{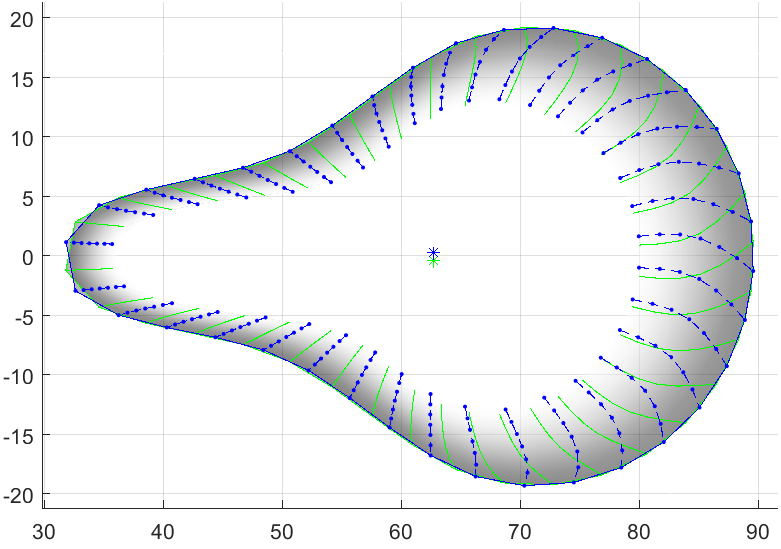}}\\
		$t=0.0165$&$t=9.0015$&$t=18.4365$
	\end{tabular}
	\caption{Chemotaxis Experiment \ref{exp:ch.c.0.5} with $c=0.5$. The stronger chemical signal from the ``east side'' leads to a higher polymerization velocity, wider lamellipodium and more effective pulling force.}\label{fig:ch.c.0.5}
\end{figure}

\begin{figure}[t]
	\centering
	\subfigure[Chemotaxis]{\includegraphics[height=10em]{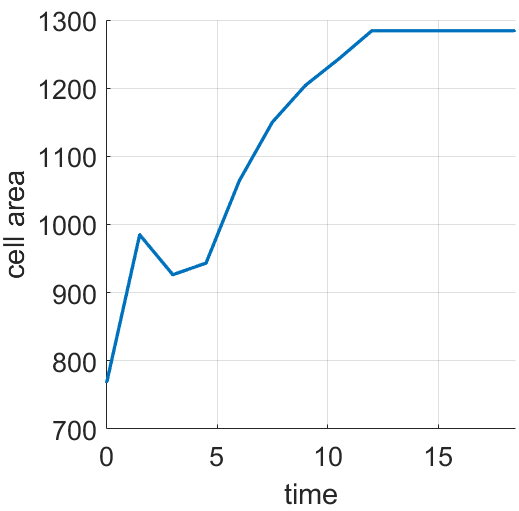}}\hspace{4em}\subfigure[Haptotaxis]{\includegraphics[height=10em]{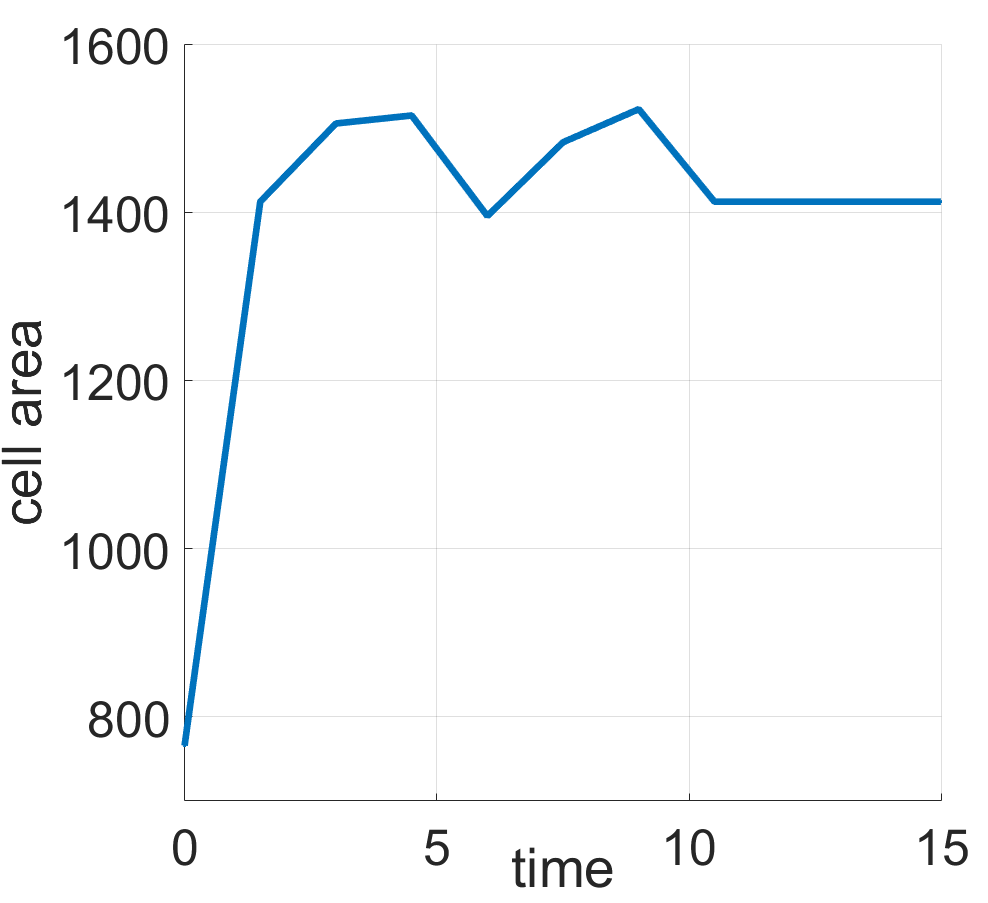}}
	\caption{Time evolution of the area the cell occupies. After an initial adjustment phase, a stable moving shape is attained. (a) Chemotaxis Experiment \ref{exp:ch.c.0.5}, (b) haptotaxis Experiment \ref{exp:hapt:lin.cont}.} \label{fig:area.evol}
\end{figure}

~
\begin{experiment}\label{exp:ch.c.0.33}
	\textbf{[More sensitive cells]:} 
	 The cell moves over a uniform substrate with $\mu^A=0.4101$. The sensitivity cut-off  value is set to $c=0.33$. The rest of the parameters are given in Table \ref{tbl:parameters}.  For this, and for the rest experiments in this paper, the resolution of the mesh is set $\snodes=7$ and $\anodes=36$. The initial conformation of the cell is assumed to be circular with a uniform-sized lamellipodium. 
	 
	 This experiment is visualized in Fig. \ref{fig:ch.c.0.33}, where we see the creation of a ``tail'' on the retracting side of the cell. After an initial transition phase, the cell continues its migration with a constant shape, see also Fig. \ref{fig:area.evol}.
\end{experiment}

~
\begin{experiment}\label{exp:ch.c.0.5}
	\textbf{[Less sensitive cells]:} With similar parameters as in Experiment \ref{exp:ch.c.0.33}, and for a sensitivity cut-off set at $c=0.5$ instead of $c=0.33$, we obtain the results exhibited in Fig. \ref{fig:ch.c.0.5}.
	
	Due to the higher value of $c$, a smaller part of the membrane senses the chemical, and the cell exhibits a longer tail  than in the $c=0.33$ case. The cell is robust and after an initial adjustment phase it attains a stable moving shape.
\end{experiment}


\section{Haptotaxis driven cell migration}\label{sec:hapt}
\begin{figure}[t]
	\centering
	\begin{tabular}{ccc}
		{\includegraphics[height=10em]{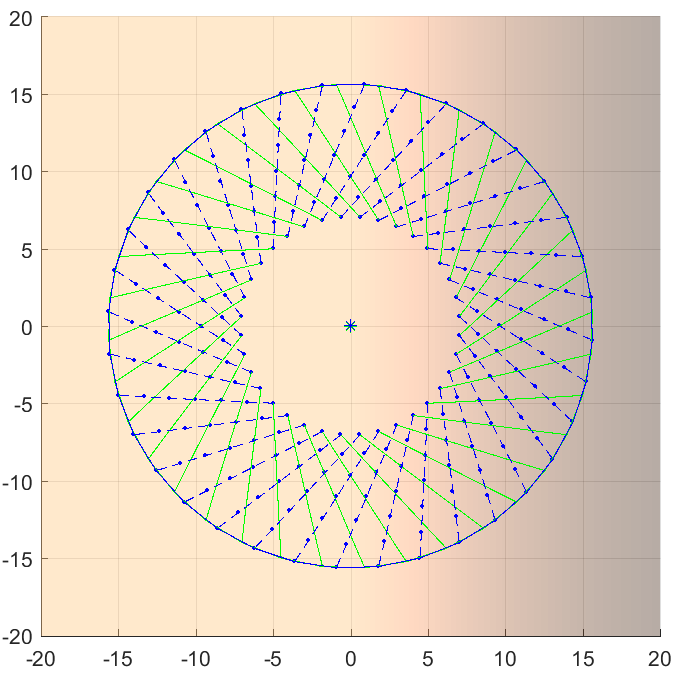}}&
		{\includegraphics[height=10em]{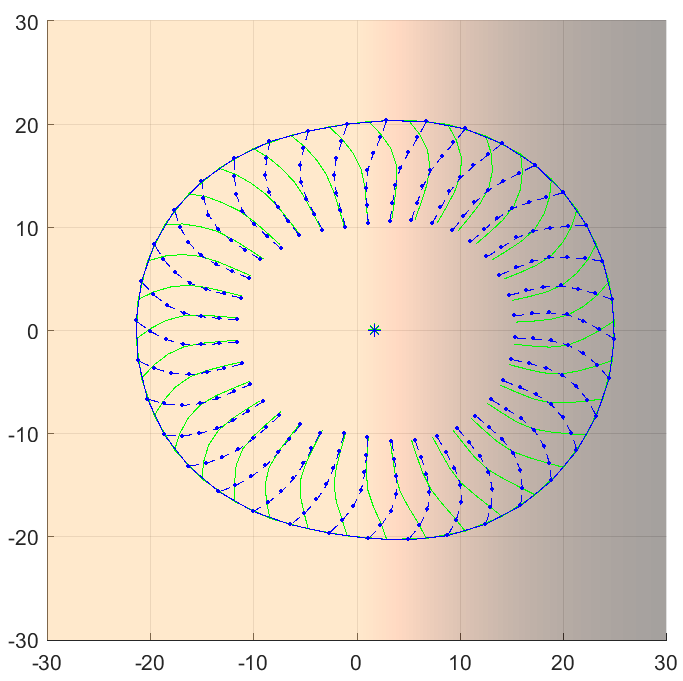}}&
		{\includegraphics[height=10em]{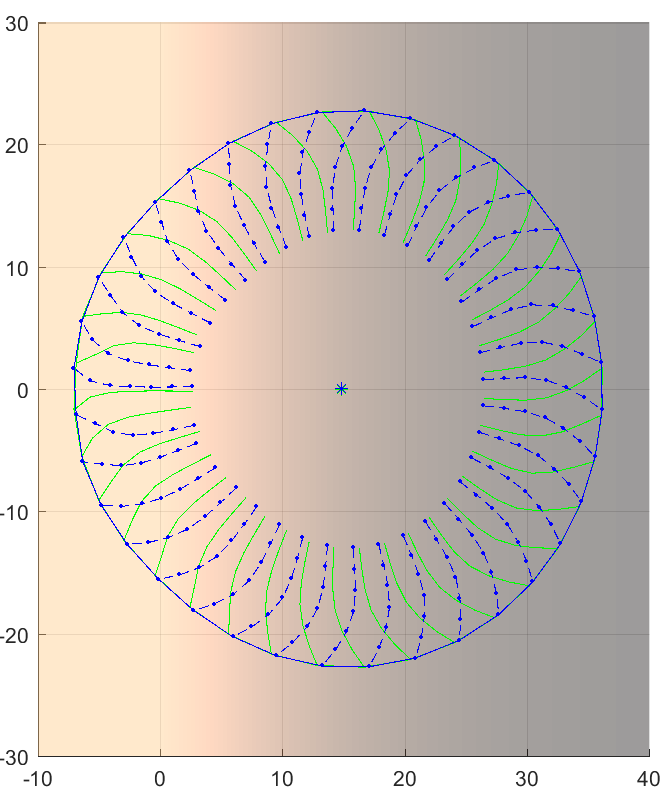} \includegraphics[height=10em]{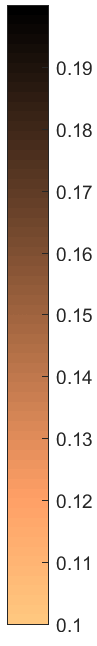}} \\
		$t=0.0015$&$t=3.0015$&$t=9.0015$
	\end{tabular}
	\caption{Continuous non-linear adhesion haptotaxis Experiment \ref{exp:hapt:nonlin.cont}. The large gradients are introduced to represent tissue in reparation. As with the Experiment \ref{exp:hapt:lin.cont}  before, the cell migrates towards the higher adhesion regions. The shape of the cell is influenced by the particular adhesion coefficient $\mu^A$.}\label{fig:exp_cont}
\end{figure}
\begin{figure}[t]
	\centering
	\begin{tabular}{ccc}
		\subfigure[$t=0.0015$]{\includegraphics[height=10em]{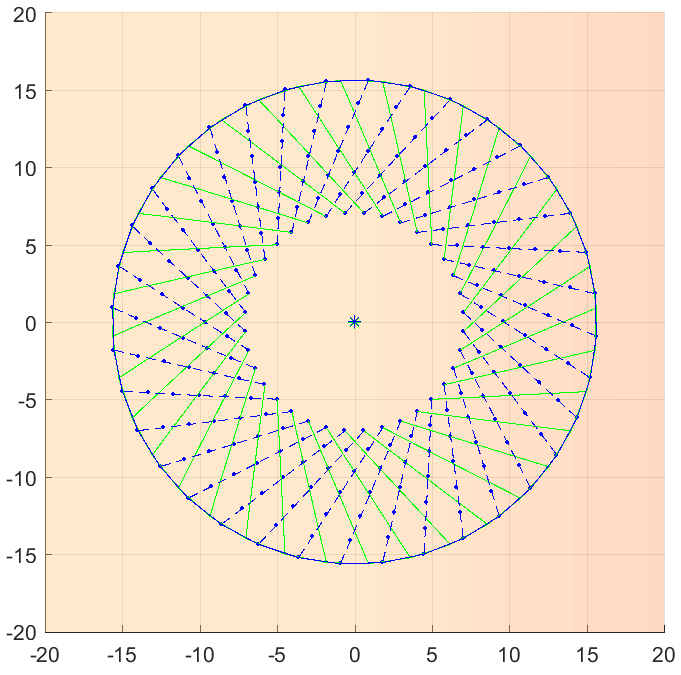}}&
		\subfigure[$t=3.0015$]{\includegraphics[height=10em]{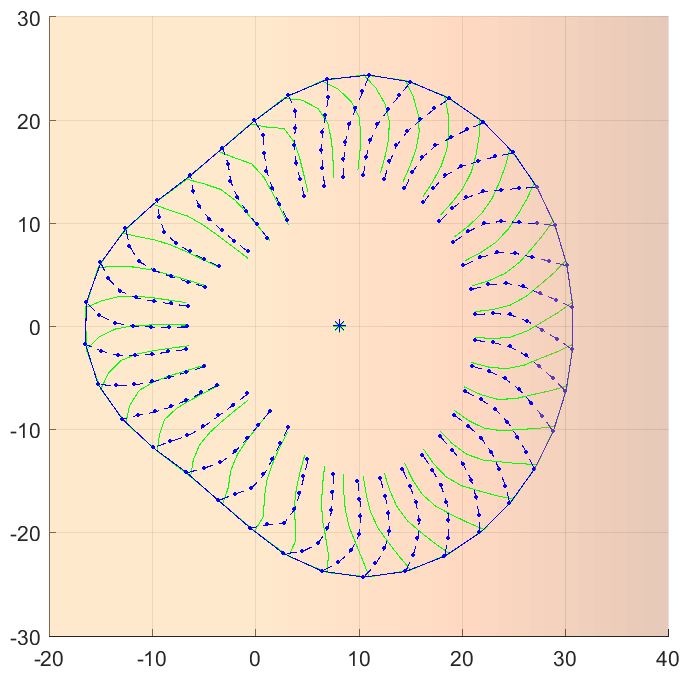}}&
		\subfigure[$t=12.0015$]{\includegraphics[height=10em]{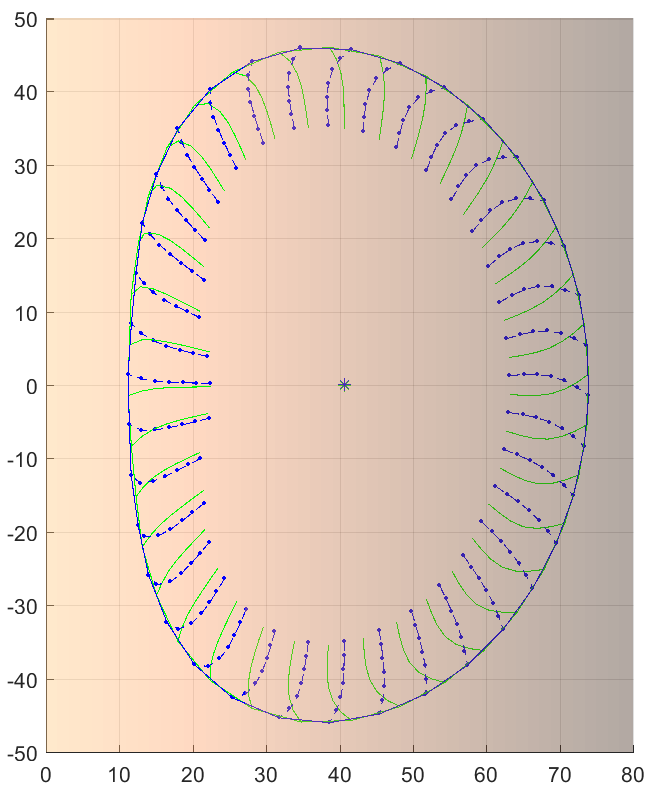}}
		\includegraphics[height=10em]{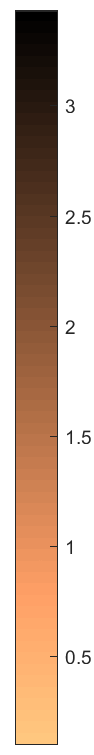}
	\end{tabular}
	\caption{Haptotaxis Experiment \ref{exp:hapt:lin.cont}: the cell resides over a non-uniform ECM with density described by the background colour and colour bar. The higher adhesion in the right side of the domain and the internal myosin pulling force lead to a larger cell which migrates to the right.}\label{fig:lin_cont}
\end{figure}

Here we study the influence that haptotaxis has on the motility of the cells, in the {absence} of a chemotaxis influence. We take haptotaxis into account by the density of the ECM fibres and the coefficient $\mu^A$ in \eqref{eq:strong}. Variations in the density of the ECM are introduced by a spatial non-uniform adhesion coefficient $\mu^A$.

The polymerization velocity $v$ is considered constant, but can be locally adjusted by the curvature of the membrane as in \eqref{eq:v.mbrn}. There is no influence on the $v$ by the density or any other characteristic of the ECM. 

Biologically, the condition of the ECM is a strong indication of health. For cancer in particular, the interaction of the cancer cells with the ECM is of fundamental importance for the invasion and the metastasis steps of the disease. For that reason we investigate three particular cases: a ``healthy tissue'' experiment where the ECM is smooth and exhibits small gradients, a ``tissue restoration'' experiment where the ECM is smooth and exhibits larger gradients, and a ``damaged tissue'' experiment where the ECM exhibits discontinuities.

~
\begin{experiment}\label{exp:hapt:nonlin.cont}
	\textbf{[{Normal tissue}]:} We consider an ECM that varies spatially in a non-linear but smooth manner as follows:
	$$\mu^A(x,y) =0.4101\times\begin{cases} 0.1, & x<0\\ \frac{0.2}{1+e^{-0.1x}}, & x\geq 0 \end{cases}.$$

	We visualize the results in Fig. \ref{fig:exp_cont} and note the smooth transition of the cell in the direction of the higher gradient. As a result of the higher adhesion, the cell at first elongates and creates an effective pulling force towards one side of the cell. We also see that, after an initial growing phase, the cell maintains a robust moving cell shape.
\end{experiment}

~
\begin{experiment}\label{exp:hapt:lin.cont}
	\textbf{[{Tissue repair}]}: During the remodelling process, the ECM exhibits continuous gradients. We model this effect by the adhesion function
	$$\mu^A(x,y) = 0.4101\times \begin{cases} 0.1, & x<0\\ 0.1 + x/30, & x\geq 0 \end{cases}.$$
		
	We visualize the result in the Fig. \ref{fig:lin_cont}, where we see that the higher gradient that the ECM presents, when compared to the Experiment \ref{exp:hapt:nonlin.cont}, causes the cell to grow larger and to create an effective pulling force towards the higher adhesion part of the domain. As a result of the internal myosin pulling force, the protruding front is followed by a retracting ``tail'' that ceases to exist after the cell has crawled on the non-constant ECM part of the domain. 
\end{experiment}

\begin{figure}[t]
	\centering
	\begin{tabular}{cccc}
		\subfigure[$t=0.0015$]{\includegraphics[height=10em]{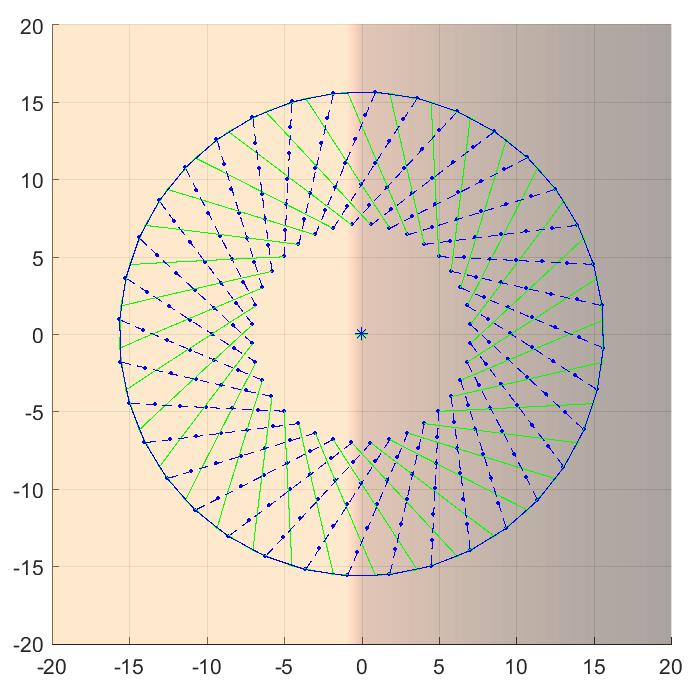}}&
		\subfigure[$t=3.0015$]{\includegraphics[height=10em]{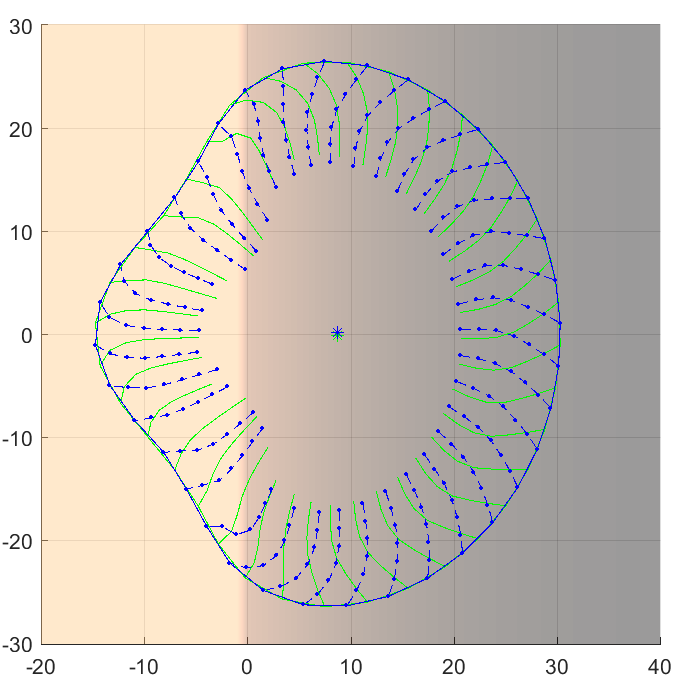}}&
		\subfigure[$t=9.0015$]{\includegraphics[height=10em]{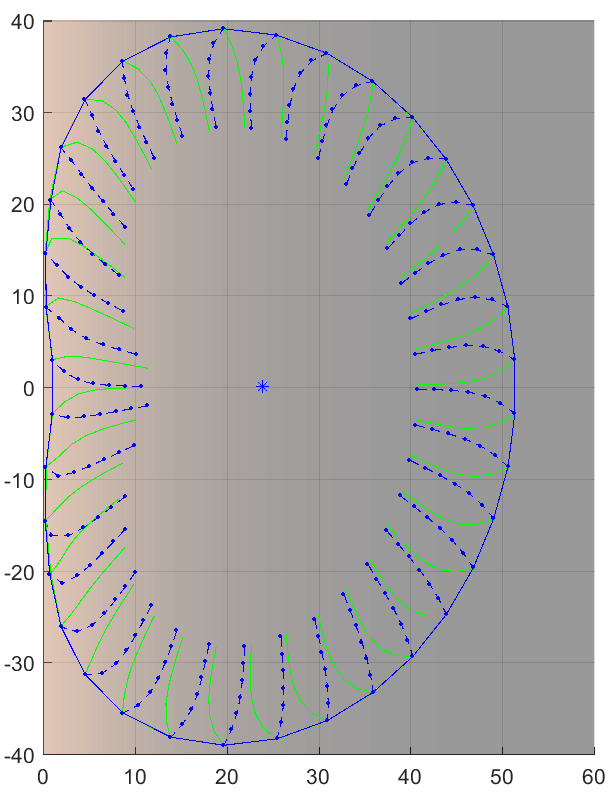}}&
		\includegraphics[height=10em]{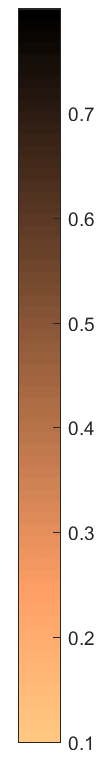}\\
	\end{tabular}
	\caption{Haptotaxis Experiment \ref{exp:hapt:nonlin.discont}, representation of a damaged tissue. When compared with Fig. \ref{fig:exp_cont} we primarily see that the cell attains a different shape during the transition to the higher adhesion substrate; influenced by the particular adhesion coefficient $\mu^A$.}\label{fig:exp_discont}
\end{figure}

~
\begin{experiment}\label{exp:hapt:nonlin.discont}
	\textbf{[{Damaged tissue}]:} Of particular medical and biological importance are cases where the ECM exhibits abrupt variations in its consistency, characteristic examples are possible wounds, or the degradation of the matrix by the cancer cells. By introducing a discontinuity in the ECM, we partially replicate these cases:
	$$\mu^A(x,y) = 0.4101\times\begin{cases} 0.1, & x<0\\ \frac{0.8}{1+e^{-0.1x}}, & x\geq 0 \end{cases}\, .$$
	
	We visualize the results in Fig. \ref{fig:exp_discont}, where we see that despite the discontinuity, the cell adheres and crawls once again to the more stiff part of the ECM. When compared with the previous Experiment \ref{exp:hapt:nonlin.cont}, we see that the ``tail'' the cell develops, is sharper despite the small gradient of the ECM. The smaller values of the ECM density are also depicted in the final size of the cell, that is small in this case. 
\end{experiment}

\section{Comparison of chemotaxis and haptotaxis on cellular level}\label{sec:comparison}

We have seen in the previous sections that due to biological properties and modelling considerations, the migration of the cells under separately chemotaxis and haptotaxis is considerably different. 

On the one hand, in chemotaxis, the wideness of the lamellipodium is influenced by the extracellular chemical signal and the polymerization velocity. The lamellipodium grows larger when the chemical signal is stronger. As a result the total friction and the effective pulling force is stronger towards this direction. On the other hand, in haptotaxis, variations of the ECM lead to larger cell sizes (not lamellipodium widths) in the parts of the cell that reside over higher ECM densities. As the friction is stronger in these parts of the cell, their resistance to the contractile myosin force is equally stronger. 

Despite these qualitative differences between chemotaxis and haptotaxis, we exhibit in this section two generic experiments exhibiting that the resulting cell shapes and cell and motilities are more similar than not. 

\begin{figure}[t]
	\centering
	\begin{tabular}{ccc}
		\subfigure[chemotaxis]{\includegraphics[height=10em]{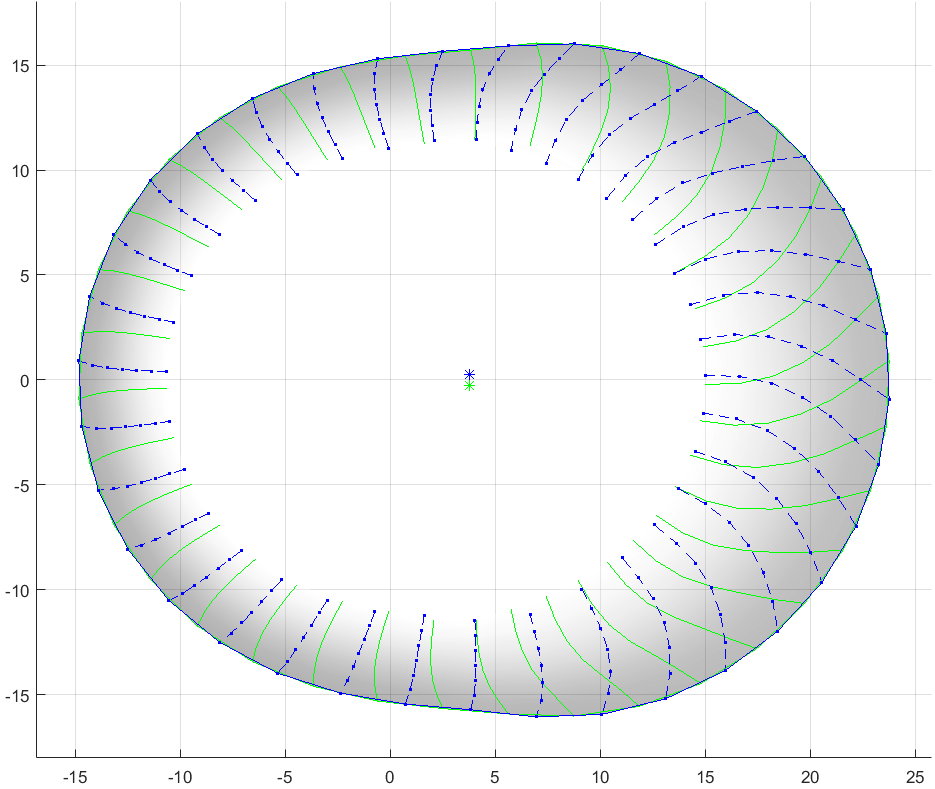}}&
		\subfigure[haptotaxis]{\includegraphics[height=10em]{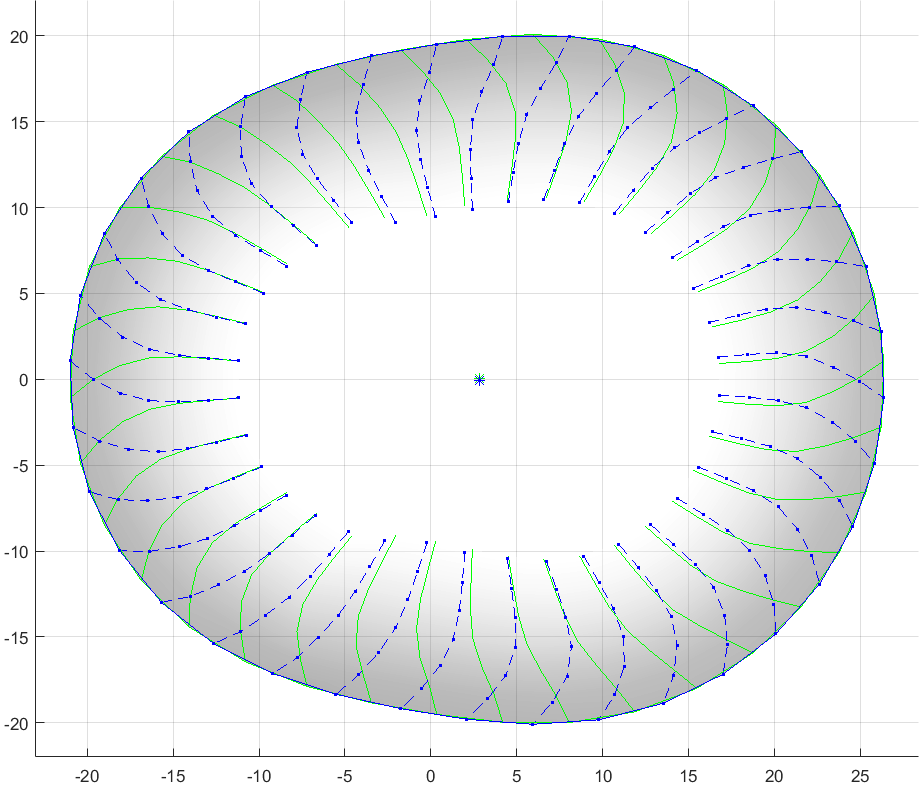}}&
		\subfigure[overlay]{\includegraphics[height=10em]{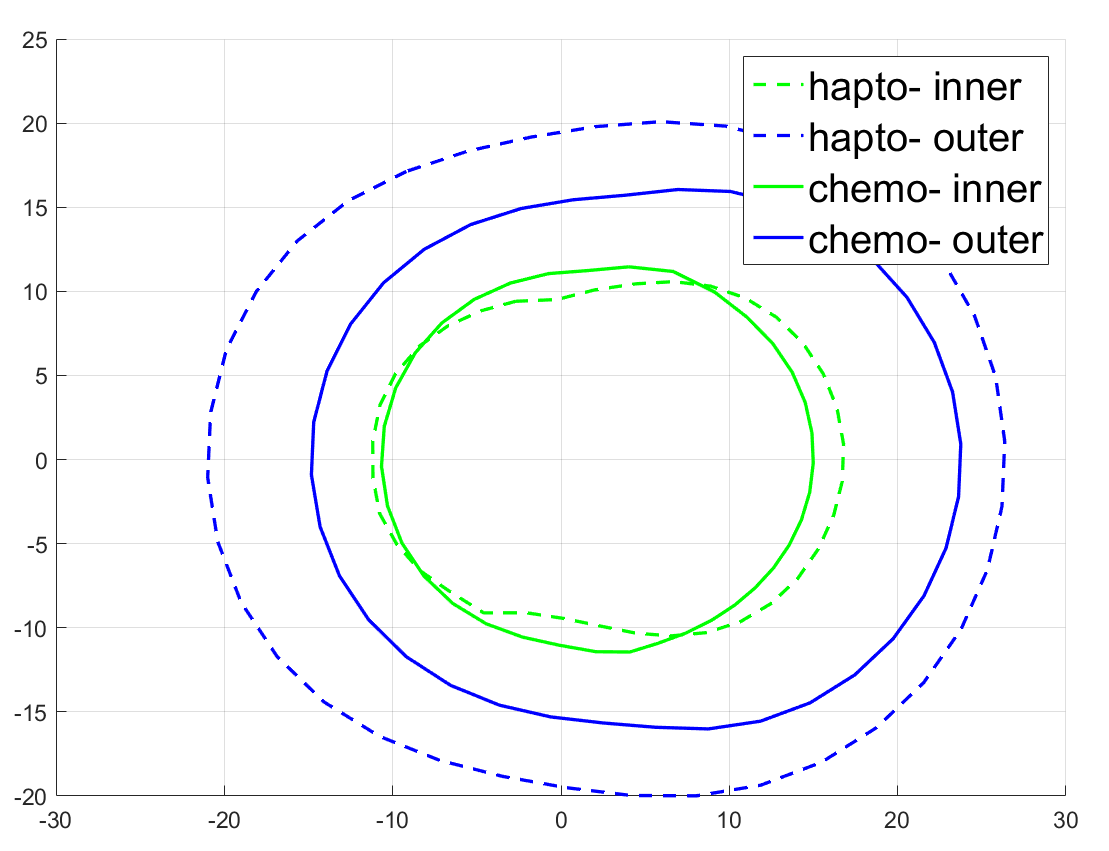}}		
	\end{tabular}
	\caption{Comparison of the chemotaxis Experiment \ref{exp:ch.c.0.5} shown in (a), with the haptotaxis Experiment \ref{exp:hapt:nonlin.cont} shown in (b).  The overlay of the outer and inner boundaries of the simulated cells, exhibits that the cells in the two cases are qualitatively similar, despite the significant biological and modelling differences between them.}\label{fig:cmp.1}
\end{figure}

~
\begin{experiment}\label{exp:cmp.1}
	For the first comparison case, visualized in Fig. \ref{fig:cmp.1}, we consider the chemotaxis Experiment \ref{exp:ch.c.0.5} and the haptotaxis Experiment \ref{exp:hapt:nonlin.cont}. We notice that the transitional conformation that the cells acquire before their final stable state, are qualitatively similar in the shape of the membrane (external boundary), but differ in size: 473 vs 461. 
	
	The size of the lamellipodium, we see in the chemotaxis case that it is not uniform whereas in the haptotaxis case seems to be the equally wide in the protruding and the retracting part of the cell. 
	
	In the ``inner boundary'' of the lamellipodium, which has more of a modelling rather than a biological interpretation, we notice similarities both in shape and size. The shape and size of this inner area, determine contractile myosin (see \cite{MOSS-model} for modelling details) and we can safely deduce that it is similar in the two cases. Any differences hence in the size and shape of the (full) cell are due to external influences.
\end{experiment}

\begin{figure}[t]
	\centering
	\begin{tabular}{ccc}
		\subfigure[chemotaxis]{\includegraphics[height=10em]{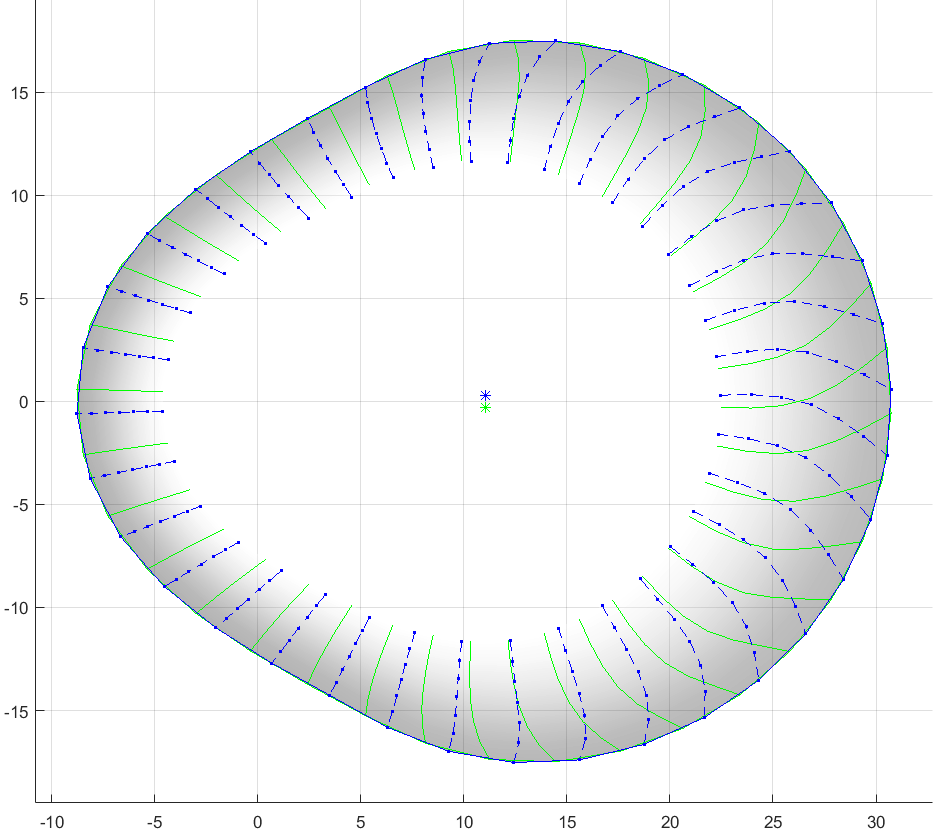}}&
		\subfigure[haptotaxis]{\includegraphics[height=10em]{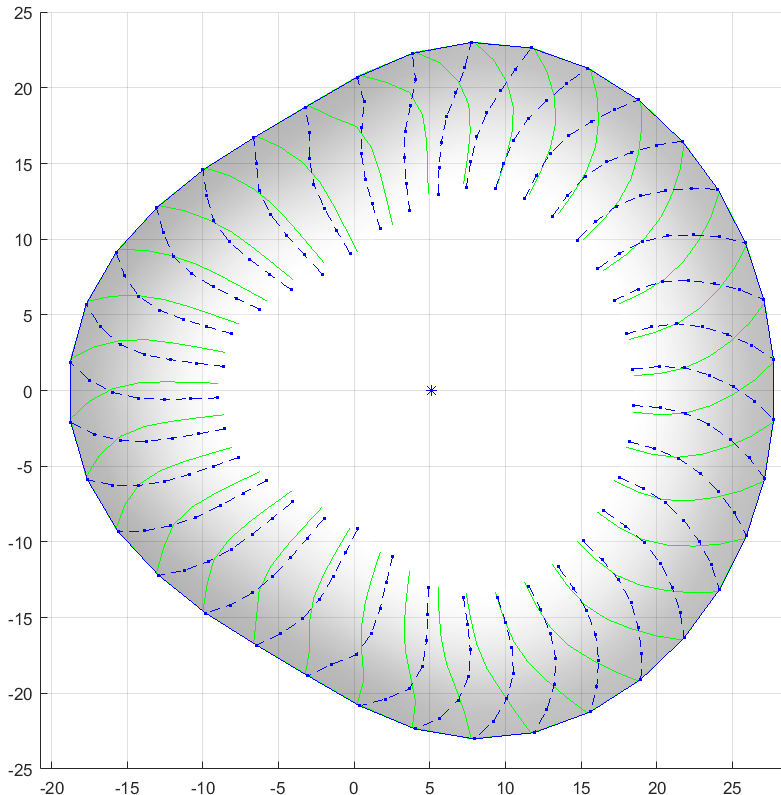}}& 
		\subfigure[overlay]{\includegraphics[height=10em]{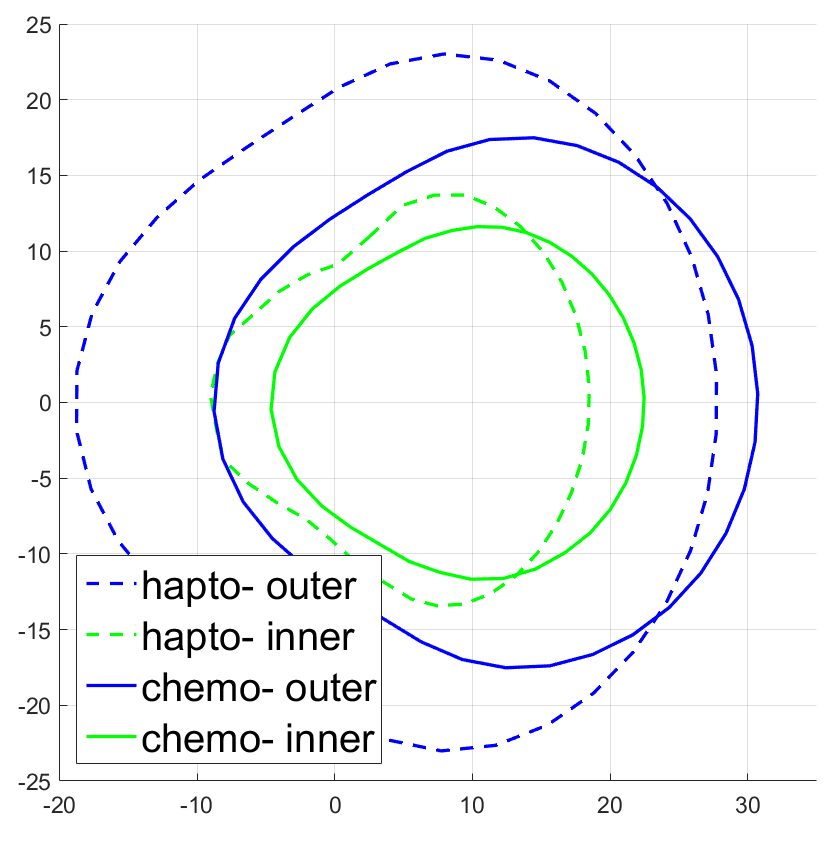}}
	\end{tabular}
	\caption{Comparison of the chemotaxis Experiment \ref{exp:ch.c.0.33} shown in (a), with the haptotaxis Experiment \ref{exp:hapt:lin.cont} shown in (b). The overlay graph shown in (c) exhibits that the areas that the cells occupy are qualitatively similar. It moreover shows that the the ``inner cells'' are similar also in size.}\label{fig:cmp.2}
\end{figure}

\begin{experiment}\label{exp:cmp.2}
	The second comparison case is visualized in Fig. \ref{fig:cmp.2} and refers to the more dynamic chemotaxis Experiment \ref{exp:ch.c.0.5} and haptotaxis Experiment \ref{exp:hapt:lin.cont}. Once again the transition shapes of the two cases are qualitatively similar despite stark differences in their biological and modelling properties. 
	
	Looking in more detail we notice further similarities between the different experimental settings. In particular, the chemotaxis experiments in Figs. \ref{fig:cmp.1} and \ref{fig:cmp.2} exhibit an inner-over-outer area ratio of approximately 0.3, where as the haptotaxis experiments an area ratio of approximately 0.45. The exact values depend on the particular parameter set that we use but they are robust within the biological accepted values, see Table \ref{tbl:parameters}.
\end{experiment}

%

\section{Discussion}
We have compared the influence that chemotaxis and haptotaxis have in the lamellipodium driven cell migration. We have used the successful and detailed FBLM to model the influences these processes have in the physiology of the cells. We have moreover employed a problem specific FEM to numerically solve the model and the motility of the cells. 

We have seen that these processes have significantly different qualitative results on the polymerization rate of the filaments, the friction forces with the substrate, the size of the cell and the lammelipodium, and on the shape of the motile cell.  

Although not exhaustive in terms of numerical experimentation, our investigation clearly exhibits that the shapes of the migrating cells are more similar than not between the two cases. 

This provides with a first indication that, an abstraction and simplification of the cell motility, could possibly merge these two effects under the same paradigm. This would facilitate the overall effort and bring the corresponding research one step closer to deriving a macroscopic model from the particular dynamical properties of single cell. 

\begin{table}[t]
	\small{
		\begin{center}
			\begin{tabular}{| c | p{0.25\linewidth} | p{0.21\linewidth} | p{0.33\linewidth} | }
				\hline
				var. & Meaning & Value & Source/Comment\\
				\hline
				& & & \\[-1em]
				$\mu^B$ & bending elasticity & $\rm 0.07\,pN\,\upmu m^2$ &\cite{Gittes1993}  \\		
				$\mu^A$ & adhesion coefficient & $\rm 0.4101\,pN\, min\, \upmu m^{-2}$& order of magnitude in \cite{Li2003,Oberhauser2002}, estimation in \cite{Oelz2008,Schmeiser2010}\\
				$\kappa_\text{br}$ & branching rate & $\rm 10\, min^{-1} $ & order of magnitude in\cite{Grimm2003}, chosen to fit $2 \overline \rho_\text{ref}=\rm 90\, \upmu m^{-1}$ \cite{Small2002}\\
				$\kappa_\text{cap}$ & capping rate & $\rm 5\, min^{-1} $ & order of magnitude in \cite{Grimm2003}, chosen to fit $2 \overline \rho_\text{ref}=\rm 90\, \upmu m^{-1}$ \cite{Small2002}\\
				$c_\text{rec}$ & Arp$2/3$ recruitment  & $\rm 900\, \upmu m^{-1}\, min^{-1} $ & chosen to fit $2 \overline \rho_\text{ref}=\rm 90\, \upmu m^{-1}$ \cite{Small2002}\\
				$\kappa_\text{sev}$ & severing rate & $\rm 0.38\, min^{-1}\, \upmu m^{-1}$ & chosen to give lamellipodium widths similar as described in \cite{Small2002} \\
				$\mu^\text{IP}$ & actin-myosin strength& $\rm 0.1\, pN\, \upmu m^{-2}$ &  \\
				$A_0$ & equilibrium inner area & $\rm 450\, \upmu m^2$ & order of magnitude as in \cite{Verkhovsky1999,Small1978}\\
				$v_\text{min}$ & minimal polym. speed & $\rm 1.5\,\upmu m\, min^{-1}$ & in biological range\\
				$v_\text{max}$ & maximal polym. speed & $\rm 8\,\upmu m\, min^{-1}$ & in biological range\\
				$\mu^P $ &pressure constant & $\rm 0.05\, pN\, \upmu m $ & \\
				$\mu^S$ & cross-link stretching & $\rm 7.1\times 10^{-3}\ pN\, min\, \upmu m^{-1} $& \\
				$\mu^T$ & cross-link twisting & $\rm 7.1\times 10^{-3}\, \upmu m$ &\\
				$\kappa_\text{ref}$ & ref. membrane curvature & $\rm (5\,\upmu m)^{-1}$ & \\
				\hline
			\end{tabular}
		\end{center}
		\caption{Parameter values employed in the current paper. For justification of the parameter values, refer to the above mentioned papers as well as to \cite{MOSS-model, MOSS-numeric}.}\label{tbl:parameters}	
	}
\end{table}
\bibliographystyle{plain} 
	\bibliography{FBLM-FEM_chem_vs_hapt}

\end{document}